\documentclass[a4paper]{article}
\usepackage{INTERSPEECH2019}
\usepackage[utf8]{inputenc}
\usepackage{amsmath,graphicx}
\usepackage{INTERSPEECH2019}
\usepackage{multirow}
\usepackage{multicol}
\usepackage{adjustbox}
\usepackage{dblfloatfix}
\usepackage{url,times,booktabs}
\usepackage{comment}
\usepackage{tabularx}
\usepackage{makecell}
\usepackage{cite}
\usepackage{enumerate}
\usepackage{amsmath,graphicx,url,times,booktabs, tabularx}
\usepackage{arydshln}
\title{Acoustic scene classification using teacher-student learning with soft-labels}
\name{Hee-Soo Heo$^*$\thanks{$^*$These authors contributed equally}, 
    Jee-weon Jung$^*$, 
    Hye-jin Shim, 
    and Ha-Jin Yu$^\dag$
        \thanks{$^\dag$ Corresponding author}\thanks{This research was supported by Basic Science Research Program
through the National Research Foundation of Korea(NRF) funded by the
Ministry of Science, ICT \& Future Planning(2017R1A2B4011609)}}
        
\address{School of Computer Science, University of Seoul, South Korea}
\email{
  zhasgone@naver.com,
  jeewon.leo.jung@gmail.com,
  shimhz6.6@gmail.com,
  hjyu@uos.ac.kr}
\begin{document}
\maketitle
\begin{abstract}
Acoustic scene classification identifies an input segment into one of the pre-defined classes using spectral information. 
The spectral information of acoustic scenes may not be mutually exclusive due to common acoustic properties across different classes, such as babble noises included in both airports and shopping malls. 
However, conventional training procedure based on one-hot labels does not consider the similarities between different acoustic scenes. 
We exploit teacher-student learning with the purpose to derive soft-labels that consider common acoustic properties among different acoustic scenes. 
In teacher-student learning, the teacher network produces soft-labels, based on which the student network is trained. 
We investigate various methods to extract soft-labels that better represent similarities across different scenes. 
Such attempts include extracting soft-labels from multiple audio segments that are defined as an identical acoustic scene. 
Experimental results demonstrate the potential of our approach, showing a classification accuracy of 77.36 \% on the DCASE 2018 task 1 validation set. 
\end{abstract}

\section{Introduction}
Acoustic scene classification (ASC) refers to a task that categorizes an input audio segment into one of the pre-defined acoustic scenes (classes). 
The detection and classification of acoustic scenes and events (DCASE) community, the leading platform for ASC and several related tasks, defines 10 acoustic scenes : airport, park, metro station, etc \cite{Mesaros2018_DCASE}. 
Such classes possess common acoustic properties, such as babble noises included in both airports and shopping malls. 

With advances in deep learning, various deep neural networks (DNNs) are primarily exploited for the ASC task, which are executed in a supervised manner using one-hot labels. 
However, in the conventional training scheme which use one-hot labels with an softmax output layer, the common properties among different classes cannot be considered. 
This is because each acoustic scene is trained to be orthogonal in the label space. 
Here, the term `orthogonal' refers to having no correlation between any other labels. 
We hypothesize that this strict training process is not efficient because it is opposite to the process of human perception. 
Humans can recognize that there is a similarity between different acoustic scenes. 
This inefficiency of the one-hot label technique can give rise to issues in most identification tasks, which would be further intensified in tasks, such as ASC, where the boundaries or definitions of the scenes are ambiguous. 
We expected that the similarity of each scene would be reflected in the soft-label. 

% soft label 하면 역시 TS지.
One of the well-known techniques that extract soft-labels is teacher-student (TS) learning. 
In the TS learning, the student network is trained using a soft-label from the teacher network rather than conventional one-hot label. 
The TS framework was first proposed for model compression, but was found to be an effective scheme to deal with various issues in related tasks, such as compensating far-field utterances in speech recognition \cite{kim2018bridgenets}.
The key factor that leads the various applications of TS learning was an appropriate modification of the framework to suit the purpose. 
Therefore, we exploited various modifications of TS learning and evaluated them to consider common properties among different classes in the ASC task. 
% In this study, we explored the applications of TS learning to the ASC task. 
% The main motivation is to validate our hypothesis that the soft-label obtained from teacher DNN can consider common properties across different classes that is not feasible using one-hot labels, as discussed in Section 1. 

With the inspiration from the previous researches \cite{jung2018short, kim2018bridgenets}, we explored two techniques to modify the TS learning to better conduct the ASC task. 
The first is extraction of soft-labels from multiple input segments. 
This method enables extracting more general soft-labels and also includes the effect of data augmentation. 
The second is direct comparison of embeddings rather than the output layer, proposed in \cite{jung2018short}, was applied for the ASC task. 
We verified that the combination of the two techniques can further improve the performance of the ASC system. 

The rest of the paper is organized as follows. 
Section 2 describes the overall system, mainly front-end DNN and back-end support vector machine (SVM), used in this study. 
The TS learning scheme and the proposed techniques are discussed in Section 3. 
Section 4 presents the experiment and analysis results, and the paper is concluded in Section 5.

\section{System description}
This section describes the system used for the ASC task. 
We use a convolutional neural network (CNN) to process the raw waveform, and a CNN with a layer of gated units (CNN-GRU) to process spectrograms. 
These two front-end models extract a fixed-dimensional embedding from an input segment. 
We used SVM as the back-end for the ensemble of the two models. 
The scheme for combining the two models using spectrograms and raw waveforms follows that of the authors' previous research in \cite{Jung2018dcase}. 
In addition, we have improved the performance of the system through few modifications, such as the DNN architecture. 
Table 1 shows that the baseline systems used in this study outperform the previously reported baselines \textbf{without the application of TS learning}.

\begin{table}[h]
\caption{Classification accuracy (\%) of individual systems and their score-sum ensemble in terms of fold 1 configuration on the validation set.}
\label{table:improvedBaseline}
\centering
\begin{adjustbox}{width=1\columnwidth}
\begin{tabular}{l c c}
\hline
\textbf{System} & \textbf{Jung \textit{et al.}} \cite{Jung2018dcase,Mesaros2018_DCASE} & \textbf{Improved baseline}\\
\hline
Raw waveform & 67.15 &  69.38 \\
Spectrogram & 66.24 & 72.73 \\
i-vector & 63.74 & - \\
\textbf{Ensemble}& 73.82 & \textbf{74.42} \\
\hline

\end{tabular}
\end{adjustbox}
\end{table}

\subsection{Extraction of CNN-GRU embedding}
State-of-the-art systems in the ASC task comprise deep architectures, using CNNs and recurrent architectures \cite{Sakashita2018, Dorfer2018, Zeinali2018, Jung2018dcase}. 
The CNN model for raw waveforms adopts 1D convolutional layers for direct processing of raw waveforms. 
Utilizing vast audio segments with a high sampling rate, these systems are capable of extracting an embedding that is highly representative. 

The DNN used in this study comprises residual convolutional blocks and a fully-connected layer similar to that of \cite{jung2018complete, jung2018avoiding}. 
In this architecture, the input segments are processed using convolutional layers to extract the frame-level features. 
These embeddings are then aggregated into an utterance-level feature by using a global max pooling layer. 
One fully-connected layer is then used to extract the embedding, followed by the output layer. 
After the training, the output layer is removed, and embeddings are extracted from the last fully-connected layer. 
Table \ref{table:waveformDNN} depicts the overall DNN architecture using raw waveforms. 

\begin{table}[t]
 \caption{DNN architecture of raw waveform model with input sequence shape: ($479999\times2$).}
  \centering
  \label{table:waveformDNN}
  \begin{adjustbox}{width=0.8\columnwidth}
  \begin{tabular}{r c c c}
  \toprule
   \textbf{Layer} & \textbf{Output shape}  & \textbf{Kernel size}  & \textbf{Stride} \\
  \hline
  Conv1 & $39999\times 64$ & $12$ &$12$\\
  \hline
  Res1 & $ 13333\times64$ & $3$ &$1$\\
  \hline
  Res2 & $ 4444\times128$ & $3$ &$1$\\
  \hline
  Res3 & $ 1481\times 128$ & $3$ &$1$\\
  \hline
  Res4 & $ 493\times 128$ & $3$ & $1$\\
  \hline
  Res5 & $ 164\times 128$ & $3$ & $1$\\
  \hline
  Res6 & $ 54\times 128$ & $3$ & $1$\\
  \hline
  Res7 & $ 18\times 128$ & $3$ & $1$\\
  \hline
  GlobalPool & $128$ & - & -\\
  \hline
  Dense1 & 64 & $128\times64$ & -\\
  \hline
  Output & 10 & $64\times10$ & -\\
  \bottomrule
  \end{tabular}
  \end{adjustbox}
\end{table}

We used a CNN-GRU model with 2D convolutional layers for processing spectrograms. 
In this model, a two-channel spectrogram extracted from stereo audio is input to the CNN, and a fixed dimensional embedding is output via the GRU layer. 
The overall DNN architecture using spectrograms is depicted in Table \ref{tab:specDNN}.

\begin{table}[h]
 \caption{DNN architecture of spectrogram model with input sequence shape: ($249\times 256 \times2$).}
  \centering
  \label{tab:specDNN}
  \begin{adjustbox}{width=0.8\columnwidth}
  \begin{tabular}{r c c c}
  \toprule
   \textbf{Layer} & \textbf{Output shape}  & \textbf{Kernel size}  & \textbf{Stride} \\
  \hline
  Conv1 & $249 \times 256 \times 30$ & $7\times 7$ &$1\times1$\\
  \hline
  Res1 & $ 249 \times256\times 30$ & $3\times 3$ &$1\times1$\\
  \hline
  Res2 & $ 125 \times128 \times 60$ & $3\times 3$ &$2\times2$\\
  \hline
  Res3 & $ 63 \times64 \times 120$ & $3\times 3$ &$2\times2$\\
  \hline
  Res4 & $ 21 \times22 \times 120$ & $3\times 3$ &$3\times3$\\
  \hline
  AvgPool & $ 21 \times 1 \times 240$ & $1 \times 22$ & $ 1 \times 22$\\
  MaxPool & $ 21 \times 1 \times 240$ & $1 \times 22$ & $ 1 \times 22$\\
  \hline
  Concan & $ 21 \times 480 $ & - & - \\
  \hline
  GRU & $ 480$ & - & -\\
  \hline
  Dense1 & 64 & $480 \times 64$ & -\\
  \hline
  Output & 10 & $64 \times 10$ & -\\
  \bottomrule
  \end{tabular}
  \end{adjustbox}
\end{table}

\subsection{SVM classification}
The DNNs with the output layer activated by the softmax function are well-known as a high-performance classifier. 
However, the softmax values in the output layer do not represent the concept of confidence. 
In other words, the softmax values are ``poorly calibrated'' \cite{guo2017calibration}. 
In case of a single DNN, this issue does not cause any problems. 
However, when combining output results from multiple DNNs, this can cause problems because the outputs are not confidence scores. 
To avoid this problem, we implemented a separate scoring phase using the SVM classifier. 
In the scoring phase, the output layers of two models are removed and the outputs of the last hidden layer of each model are trained using one SVM for each model. 
Finally, the scores calculated from the two SVMs are averaged for the ensemble of the raw waveform and spectrogram-based models. 

\section{Teacher-student learning in ASC}
Teacher-student (TS) learning is a framework that adopts two DNNs. 
In this scheme, a superior system (teacher DNN) is first trained using conventional training scheme with one-hot labels. 
Superiority of the teacher DNN is determined depending on the target task, i.e., larger capacity for model compression. 
Then the output of the teacher DNN (referred to as soft-label) is used to train a student DNN \cite{li2014learning}. 
Specifically, the output distributions of the teacher and student DNNs are compared using Kullback-Leibler divergence with the following equation: 
\begin{equation}
  \label{eqn:short utterance ts learning}
    \displaystyle TS_{org} = - \sum\limits_i^I\sum\limits_j^Jp_T(o_j|x_{i}) \log(p_S(o_j|x_{i})),
\end{equation}
where $p_T(\cdot)$ and $p_S(\cdot)$ are the output distributions of the teacher and student network, respectively, and $x_{i}$ refers to the $i'th$ input. 
%Eq. (1) is not the exact KL-divergence, but the same effect can be achieved when training the student network.
Eq. (1) refers to cross-entropy rather than KL-divergence, but the same effect can be achieved when training the student network (look \cite{li2014learning}, Section 3.1. for further details). 

The TS framework has been expanded to knowledge distillation with the concept of temperature in \cite{hinton2015distilling}.
In knowledge distillation, the temperature $T$ adjusts the extent of soft-label utilization where a higher $T$ softens the probability distribution. 
Hence, the original equation of TS learning is expanded by including temperature variable $T$ as:
\begin{equation}
  \label{eqn:short utterance ts learning}
    \displaystyle TS_{T} = - \sum\limits_i^I\sum\limits_j^Jp_T(o_j|x_{i};T) \log(p_S(o_j|x_{i})),
\end{equation}
\begin{equation}
  \label{eqn:short utterance ts learning}
    \displaystyle p_T(o_j|x_{i};T) = \frac{exp(p_T(o_j|x_{i})/T)}{\Sigma_{k}^{K}exp(p_T(o_k|x_{i})/T)} ,
\end{equation}

% The TS framework was first proposed for model compression, but was found to be an effective scheme for various tasks, such as compensating far-field utterances. 
% In this study, we explored the applications of TS learning to the ASC task. 
% The main motivation is to validate our hypothesis that the soft-label obtained from teacher DNN can consider common properties across different classes that is not feasible using one-hot labels, as discussed in Section 1. 

\begin{figure}[t!]
  \centering
  \includegraphics[width=0.8\columnwidth]{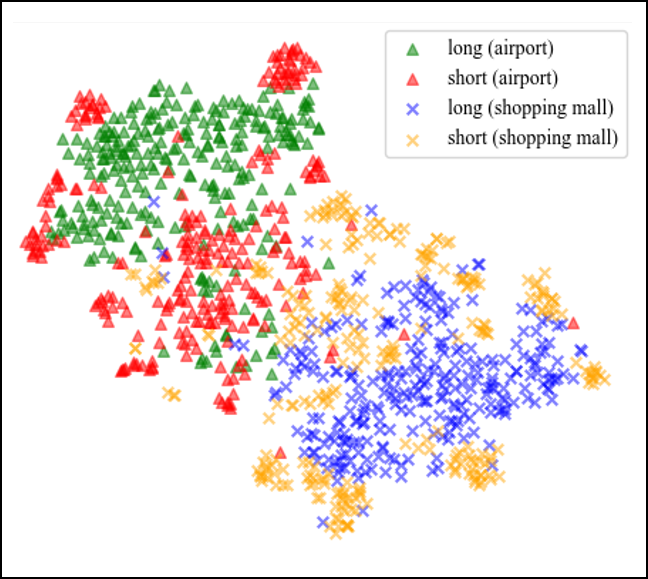}
  \caption{Illustration of the superiority using multiple segments for longer duration for the teacher network. t-SNE \cite{maaten2008visualizing} plot of embeddings extracted from test set segments on the most confusing acoustic scene pair, `shopping\_mall-airport'. $\bigtriangleup$ and $\times$ symbols represent the embeddings of shopping\_mall and airport scene, respectively. Cohesion of long duration embeddings are stronger, demonstrating the superiority of teacher network's input.} 
  \label{fig:1}
\end{figure}

\subsection{Concatenating multiple inputs}
One of the important issues in the TS learning scheme is to design the superiority of the teacher network. 
In the study that proposed the TS scheme, large capacity of the teacher network was the superiority for a model compression task \cite{li2014learning}. 
For far-field compensation study, less noise and reverberation comprised the superiority where the utterances recorded from close talk and the utterances recorded from far-field are input to the teacher and the student respectively \cite{li2018developing, kim2018bridgenets}. 
In our study, possession of additional recording from identical acoustic scenes is set as the superiority. 
For this superiority, additional recording from the same class is concatenated to every utterance, and then used for extracting the soft-label. 
For instance, the soft-label of a student DNN with input utterance `A' will be derived by concatenating another utterance `B' from the same class and then inputting to the teacher DNN. 
Here, utterance `A' is input to the student DNN and concatenation of utterance `A', `B' is input to the teacher DNN to extract a soft-label. 
This scheme is represented using the following equation: 
\begin{equation}
  \label{eqn:short utterance ts learning}
    \displaystyle KL_{loss} = - \sum\limits_i^I\sum\limits_j^Jp_T(o_j|x_{i, con}) \log(p_S(o_j|x_{i, base})),
\end{equation}
\noindent where $x_{i,base}$ is the base segment and the input segment of the student network, and $x_{i,con}$ is the input segment of the teacher network constructed by concatenating the input segment of the student DNN with another segment from an identical class. 
In particular, the length of $x_{i,base}$ is fixed to 10 s, and the duration of $x_{i,con}$ can be 20, 30 s, or longer. 
The superiority caused by concatenating multiple inputs is shown in detail in figure 1. 
The figure shows that the embeddings extracted from long segments have higher discriminative power. 
This superiority is also interpreted as that of short utterance compensation in the field of speaker verification \cite{jung2018short, Yamamoto}. 

Deriving soft-labels using multiple audio recordings is also expected to include the effect of data augmentation. 
Training data augmentation, conducted by adding noises or shifting pitch, is a well-known technique for performance improvements in the audio processing tasks using DNNs. 
In the ASC task, however, it is difficult to apply the conventional data augmentation scheme because the definition of acoustic scene is ambiguous, and there is no clear approach to define which noise belongs to which class. 
In particular, adding babble noise to the training data can give rise to a critical issue that changes the class label of the training data. 
In the proposed approach, the input data of the teacher network is changed depending on the selection of inputs for concatenation where the input audio recording of the student network is fixed. 
Therefore, we expected that there would be a similar effect of data augmentation through these combinations of multiple inputs.

\subsection{Learning based on embeddings distance}
In the proposed ASC system introduced in Section 2, the softmax output layer is removed after the training phase and the embeddings extracted from the last hidden layer are used for the scoring phase. 
Therefore, the last hidden layer, and not the output layer, is a more dominant factor in the performance of the ASC system. 
Based on this property, we modified the TS learning to take into account the output of the last hidden layer as follows: 
\begin{equation}
  \label{eqn:short utterance ts learning}
    \displaystyle TS_{emb} = \sum\limits_j^J Dist(E_T(x_{j}), E_S(x_{j})),
\end{equation}

\noindent where $Dist(\cdot)$ is the distance measure between two vectors, and $E_T(\cdot)$ and $E_S(\cdot)$ are the output of the last hidden layer from the teacher and the student network, respectively.
In this study, we used mean squared error as the distance measure. 
We interpreted this approach as the distilling the knowledge at the last hidden layer, and not at the output layer. 
This approach was inspired by \cite{jung2018short} and \cite{kim2018bridgenets}. 

\section{Experiments and results}
In this section, we show the results of experiments to evaluate the various TS learning techniques. 
The baseline system used for the performance comparison is an improved version of the authors' system that was presented at the last DCASE 2018 competition (see Table 1). 
Therefore, we focused on evaluating the effectiveness of the TS learning rather than comparing it with other systems. 

\subsection{Dataset}
DCASE 2018 task 1-a dataset \cite{Mesaros2018_DCASE} was used for all experiments in this study. 
This dataset comprises 864 audio segments that has 10 s duration for each of 10 pre-defined classes, resulting in a total of 8640 segments. 
The audio segments were recorded in stereo at a sampling rate of 48 kHz. 
Cross-validation was conducted using the four fold configuration provided within the DCASE dataset, where the validation sets are recorded from different locations. 

\subsection{Experimental settings}
All experiments in this study were conducted using Keras, a deep learning library for python, with Tensorflow back-end \cite{keras, tensorflow, tensorflow2}. 

For the raw waveform model, pre-emphasis is applied \cite{Pre-emphasis} without any other pre-processing. 
The whole segment is input, which makes the shape of input segment as $(479999, 2)$. 
The DNN architecture configuration is depicted in Table \ref{table:waveformDNN}.

We extracted the spectrograms of 256 coefficients from 100 ms windows for every 40 ms. 
The spectrogram of $(249, 256, 2)$ shape was extracted for each segment that contains stereo audio of 10 s. 

Adam optimizer \cite{kingma2014adam} with 0.001 learning rate was used for training both CNN using raw waveforms and CNN-GRU using spectrograms. 
The batch size for training the two models was 40. 
For efficient training of the spectrogram-based CNN-GRU model, we trained the CNN part except the GRU layer in the whole model and then re-trained after attaching the GRU layer on the CNN, following the multi-step training scheme reported in \cite{jung2018avoiding, E2E_Heesoo}.

\begin{figure}[t!]
  \centering
  \includegraphics[width=\columnwidth]{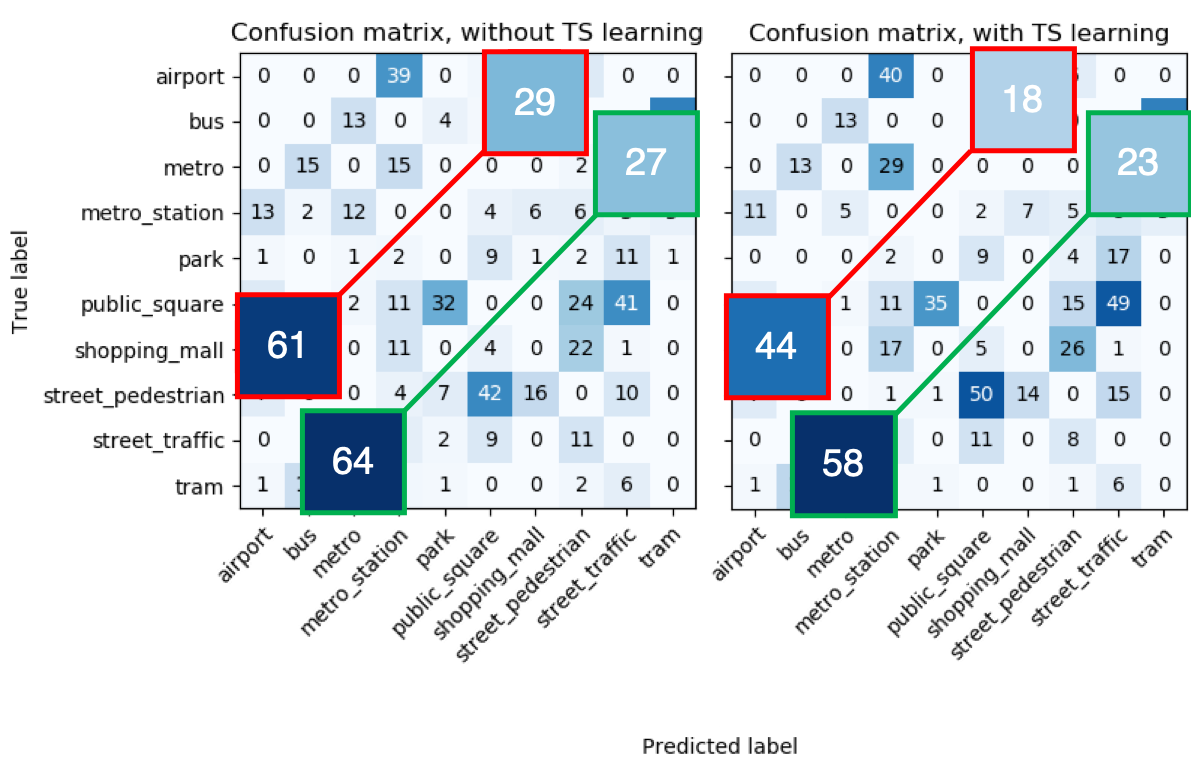}
  \caption{Confusion matrices before (\textit{left}) and after (\textit{right}) applying the proposed TS learning scheme. Both of the two most confusing acoustic scene pairs, `shopping\_mall-airport' (\textit{red box}) and `metro-tram' (\textit{green box}) show improvement.}
  \label{fig:conmat}
\end{figure}

\subsection{Analysis of results}
Table \ref{table:improvedBaseline} demonstrates the baseline of this study, which is an improved version of the authors' submission to DCASE 2018 competition. 
The individual systems show improved performance. 
For the ensemble, ``Improved baseline'' outperforms the previous baseline despite the i-vector \cite{dehak2011front} system is excluded. 

Table \ref{table:variousTemperature} shows the performances of spectrogram-based models using TS learning with various temperature $T$ and input durations. 
The temperature coefficient $T$, described in equation (2), is an important factor that determines the extent of soft-label utilization.
As the value of $T$ increases, the output distribution of the teacher network becomes more noisy. 
On the contrary, decrease in the value of $T$ refers to imposing more weight to the one-hot (hard) label. 
Comparing the results when $T$ values is one and five, we interpret that to some extent, considering common properties is important.

It is also important to design the superiority of the teacher network in the TS learning scheme. 
This is because the training process of the student network is totally dependent on the teacher network. 
However, the teacher network, trained using one-hot labels, may have poor performance. 
Even the teacher network may distill wrong knowledge to the student network.  
Experimental results showed that adjusting the input length of the teacher network to gain superiority could lead to performance improvements as intended. 

\begin{table}[h]
\caption{Performance in terms of accuracy (\%) depending on temperature $T$ and superiority of the teacher network.}
\label{table:variousTemperature}
\centering
\begin{adjustbox}{width=0.7\columnwidth}
\begin{tabular}{c|c|cc}
\hline
\multicolumn{2}{c}{\multirow{2}{*}{}} & \multicolumn{2}{c}{Duration of teacher input} \\
\cline{3-4}
\multicolumn{2}{c}{} & 10 seconds & 20 seconds  \\
\hline
\multirow{3}{*}{$T$} & 1 & 70.96 & 71.43  \\
 & 5 & 72.63 & \textbf{73.23}  \\
 & 10 & 72.51 & 72.79  \\
 \hline
\end{tabular}
\end{adjustbox}
\end{table}

Table \ref{table:knowledgeDistilliation} demonstrates the effectiveness of direct comparison of the embeddings between the teacher and student networks. 
The results show that use of the embedding of teacher DNN outperforms the soft-label of the output layer. 
Use of both soft-label and embedding layer, however, shows decreased accuracy. 
We interpret that this phenomenon occurred because the common properties across different classes are better represented in the embedding space rather than using human defined one-hot labels. 
For example, the soft-label generated at the output layer represents the common properties, such as babble noise depending on the corresponding classes (airport or shopping mall), but the soft-label at the last hidden layer can represent the common property by manifold in a high dimensional embedding space. 

\begin{table}[h]
\caption{Performance depending on points of knowledge distillation.}
\centering
\label{table:knowledgeDistilliation}
\begin{adjustbox}{width=0.8\columnwidth}
\begin{tabular}{l c}
\hline
Point & accuracy (\%) \\
\hline
output layer & 73.23 \\
output \& last hidden layer & 73.19 \\
last hidden layer & \textbf{74.26} \\
\hline

\end{tabular}
\end{adjustbox}
\end{table}

Table \ref{table:finalResult} shows the comparison of the approaches used in this study to the baseline.
The proposed approach with TS learning demonstrates classification accuracy of 77.36 \% compared to 74.42 \% for the scheme without TS learning. 
From these results, we conclude that the TS learning scheme is effective for the ASC task, and the approaches developed in this study are also valid. 

\begin{table}[h]
\caption{Classification accuracy of individual systems and their score-sum ensemble in terms of fold 1 configuration on the validation set (W/O TS: systems trained without TS learning, W/ TS: systems trained with TS learning). }
\label{table:finalResult}
\centering
\begin{adjustbox}{width=0.7\columnwidth}
\begin{tabular}{l c c}
\hline
\textbf{System} & \textbf{W/O TS} & \textbf{W/ TS}\\
\hline
Raw waveform & 69.38 & 72.99 \\
Spectrogram & 72.59 & 74.26 \\
\textbf{Ensemble}& 74.42 & \textbf{77.36} \\
\hline

\end{tabular}
\end{adjustbox}
\end{table}

We analyzed the detailed contribution of performance enhancement by TS learning.
Figure \ref{fig:conmat} illustrates two confusion matrices, with and without applying the proposed TS learning scheme, which shows that TS learning significantly reduced the errors between the two most confusing acoustic scene pairs.

\section{Conclusions and discussion}
The conventional training procedures using one-hot label cannot represent common properties among different classes. 
We assume that this scheme is not appropriate for tasks such as the ASC where the decision boundary of each class is ambiguous. 
Therefore, we explored various applications of TS learning to use the soft-label instead of the one-hot label. 
We applied TS learning to the ASC task for the first time based on two techniques: using multiple segments for the teacher network and distilling the knowledge at the last hidden layer. 
In TS learning, the student network is trained using soft-labels extracted from the teacher network. 
Soft-label in TS learning was interpreted to incorporate the correlation of different acoustic scenes with common acoustic properties. 
We evaluated various approaches of TS learning using the DCASE 2018 task 1 dataset. 
Experimental results demonstrate that designing the superiority of the teacher network and adjusting the point of knowledge distillation could improve the performance. 
In particular, TS learning significantly reduced the errors between the most confusing scenes. 

\newpage\newpage
\bibliographystyle{IEEEtran}
\bibliography{mybib}

\end{document}